\documentclass[aps,prb,preprint,groupedaddress,superscriptaddress,showkeys,
showpacs]{revtex4}%
\usepackage{graphics}
\usepackage{dcolumn}
\usepackage{bm}%
\usepackage{amsmath}%
\usepackage{amsfonts}%
\usepackage{amssymb}%
\usepackage{graphicx}

\begin{document}

\title{Knot undulator to generate linearly polarized photons with low on-axis power density}
\author{S. Qiao}
\email{qiaoshan63@gmail.com}
\affiliation{Department of Physics, Applied Physics and Stanford Synchrotron Radiation Laboratory, Stanford University, Stanford, CA 94305, USA}
\affiliation{Advanced Light Source, Lawrence Berkeley National Lab., Berkeley, CA 94720, USA}
\affiliation{Advanced Materials Laboratory, Fudan University, Shanghai 200433, China}
\author{Dewei Ma}
\affiliation{Advanced Materials Laboratory, Fudan University, Shanghai 200433, China}
\author{Donglai Feng}
\affiliation{Advanced Materials Laboratory, Fudan University, Shanghai 200433, China}
\author{Z. Hussain}
\affiliation{Advanced Light Source, Lawrence Berkeley National Lab., Berkeley, CA 94720, USA}
\author{Z. -X. Shen}
\affiliation{Department of Physics, Applied Physics and Stanford Synchrotron Radiation Laboratory, Stanford University, Stanford, CA 94305, USA}
\date{\today}

\begin{abstract}
Heat load on beamline optics is a serious problem to generate pure linearly 
polarized photons in the third generation synchrotron radiation facilities. For 
permanent magnet undulators, this problem can be overcome by a figure-8 operating 
mode. But there is still no good method to tackle this problem for electromagnetic 
elliptical undulators. Here, a novel operating mode is suggested, which can generate 
pure linearly polarized photons with very low on-axis heat load. Also the available 
minimum photon energy of linearly polarized photons can be extended much by this 
method. 
\end{abstract}
\maketitle
\section{\label{sec1}Introduction}
Undulators are broadly used in the third generation synchrotron radiation facilities. To generate low energy photons in a high energy storage ring, strong magnetic field (high K value) is necessary, which results in two serious problems: high harmonics contamination and high heat load on optical elements of the beamline. Quasi-periodic undulators\cite{HashimotoNIM,ChavanneEPAC98,DiviaccoEPAC98} have been broadly used recently to suppress the high harmonics. For helical or elliptical undulators, because the electron velocity deviates from the undulator axis, the heat load on axis is very low compared with the planar undulators and the heat load problem is only critical for the latter when pure linearly polarized photons are needed. Many novel undulators have been proposed to generate linearly polarized photons with low on axis heat load. The first solution is the figure-8 undulator suggested by T. Tanaka and H. Kitamura\cite{TanakaNIM}. In their invention, the period of magnetic field in vertical direction is two times as large as that of horizontal one, and the electrons moves in a figure-8 orbit. The velocity of electrons always deviates from the undulator axis and the electrons move through right-handed and left-handed circles alternately which results in the cancellation of circular polarization of photons and therefore only a linear polarization is remnant. The disadvantage of original figure-8 undulator is its nonability to generate circularly polarized photons. To overcome this problem, asymmetric figure-8\cite{TanakaNIM2,TanakaRSI}, and APPLE-8 undulators\cite{SasakiEPAC98} were proposed. 
However, the switching from circular to figure-8 mode can only be performed for permanent or hybrid\cite{TanakaRSI} undulators until now because the change of phase difference between vertical and horizontal magnetic field by the shift of permanent magnets along the undulator axis is crucial. Unfortunately, to generate low energy photons in a high energy storage ring, the period of undulator must be long enough and electromagnetic elliptically polarized undulator (EPU) is a more suitable choice. Because the long poles are difficult to be shifted along the undulator axis, the figure-8 operating mode can not be achieved easily due to the reason which will be discussed in the next section, and the high heat load when working in linear mode is still a serious problem for electromagnetic EPUs. 

Another possibility to resolve the heat load problem is to utilize crossed EPU\cite{KimNIM}. In this case, the cancellation of circular polarization of photons from two successive EPU, one working in right-handed polarization mode and another in left-handed mode, will generate linearly polarized photons. But in this case, the degree of polarization is strongly decreased as

$\Delta p \propto N^{2}\gamma ^{4},$

where N and $\gamma$ are the period number of EPU and the electron beam energy, respectively, so it is difficult to use crossed EPU to generate pure linearly polarized photons at high energy storage rings as pointed out by Sasaki\cite{SasakiPAC97}.

A beamline with photon energy ranging from 7 to 70 eV is planned to be built in Shanghai Synchrotron Radiation Facility by Fudan University. Because the electron beam energy of the storage ring is as high as 3.5 GeV, the heat load problem must be considered carefully. Here, we show a new operation mode of electromagnetic EPU which can generate linearly polarized photons with very low on-axis power density.

\section{Principle and Structure} 
\begin{figure}[t]
\includegraphics[width=12cm] {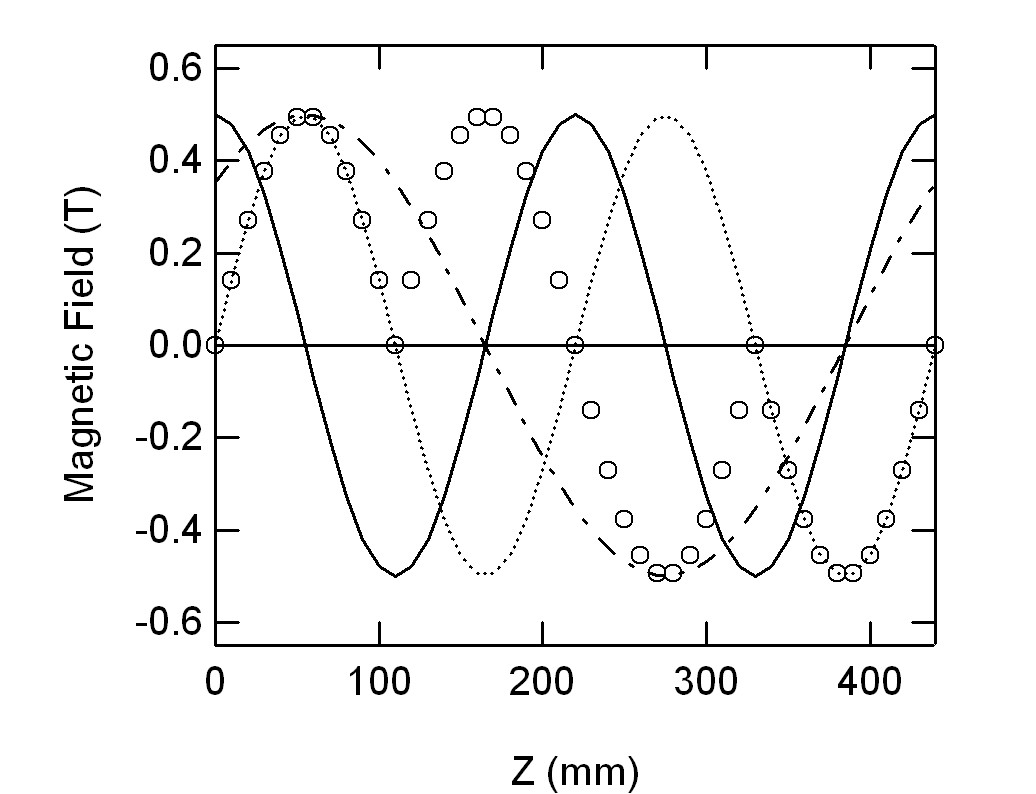}
\caption{The comparison of different magnetic fields. Solid line and broken lines are the magnetic fields in Y and X directions of a normal EPU with a 220 mm period. Circles are the magnetic field along X direction with a period of 440 mm by inverting the polarization of half poles. Solid and dash lines are the magnetic fields in Y and X directions of a figure-8 undulator.}
\label{fig1}
\end{figure}

The characteristic of figure-8 undulator is the different periods of magnetic fields in vertical and horizontal directions. The period of electromagnetic undulators can be changed easily by inversion of the polarization of some poles, so it seems that the figure-8 operation mode may be achieved easily for elliptical electromagnetic undulators. In fact, this guess is wrong and the reason can be understood from Fig.\ref{fig1}. Hereafter, the X, Y, and Z directions are defined as horizontal, vertical and undulator axial directions. The magnetic field of figure-8 undulator along X and Y directions are shown by dash and solid lines. We can see the center of gravity of the positive (negative) magnetic field in X and Y directions, while the velocity of electrons is zero, are always not coincident at the same Z position, so the electron velocity always deviates from the undulator axis. For electromagnetic EPU, after doubling the period of magnetic field in X direction by inversion of half poles, we can see that the centers of gravity of positive (negative) magnetic field in X and Y directions (circles and solid lines in Fig.\ref{fig1}) are always coincident at the same Z point where the velocities in both X, Y directions are zero. This is the reason why we can not suppress the on axis power density by only double the period of magnetic field in one direction for an electromagnetic EPU. The shift of magnets along undulator axis is also crucial here if we hope to get a figure-8 movement of electrons.

\begin{figure}[t]
\includegraphics[width=12cm] {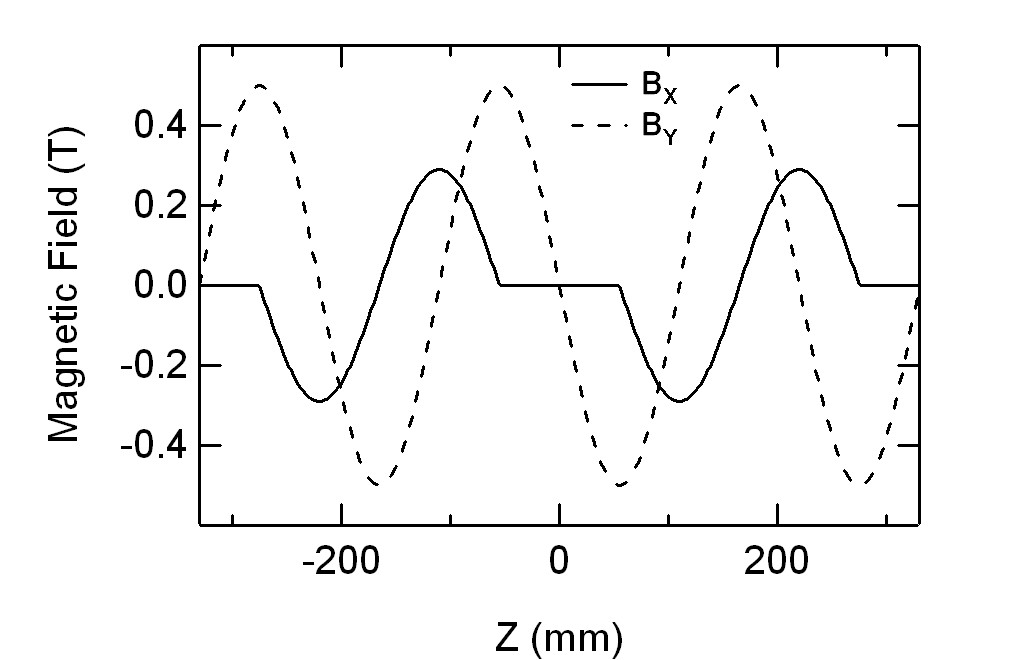}
\caption{The magnetic fields of knot undulator. The amplitudes of vertical and horizontal magnetic fields are 0.50 and 0.29 Tesla.}
\label{fig2}
\end{figure}

The period and total length of our EPU is 220 mm and about 4500 mm, respectively. The magnetic field structure of the knot undulator is shown in Fig.\ref{fig2}. The idea is very simple and just similar to that of crossed EPU. Our calculations show that the linear polarization is only 47\% when we use full length crossed EPU setup, 10 periods for both the left- and right-handed sections. Because the decrease of polarization is proportional to the square of period number as we discussed before, pure linearly polarized photons are possible to be produced by successive short crossed EPU. In the knot undulator, all the left- and right-handed EPU is only one period long. The vertical magnetic field is the same as normal EPU. The polarizations of half magnetic poles of horizontal magnets are inverted and the poles between the one period right- and left-handed EPU are turned off for the $\pi$ phase shift. The period of horizontal magnetic field is 1.5 times as long as that of vertical one.

\begin{figure}[t]
\includegraphics[width=12cm] {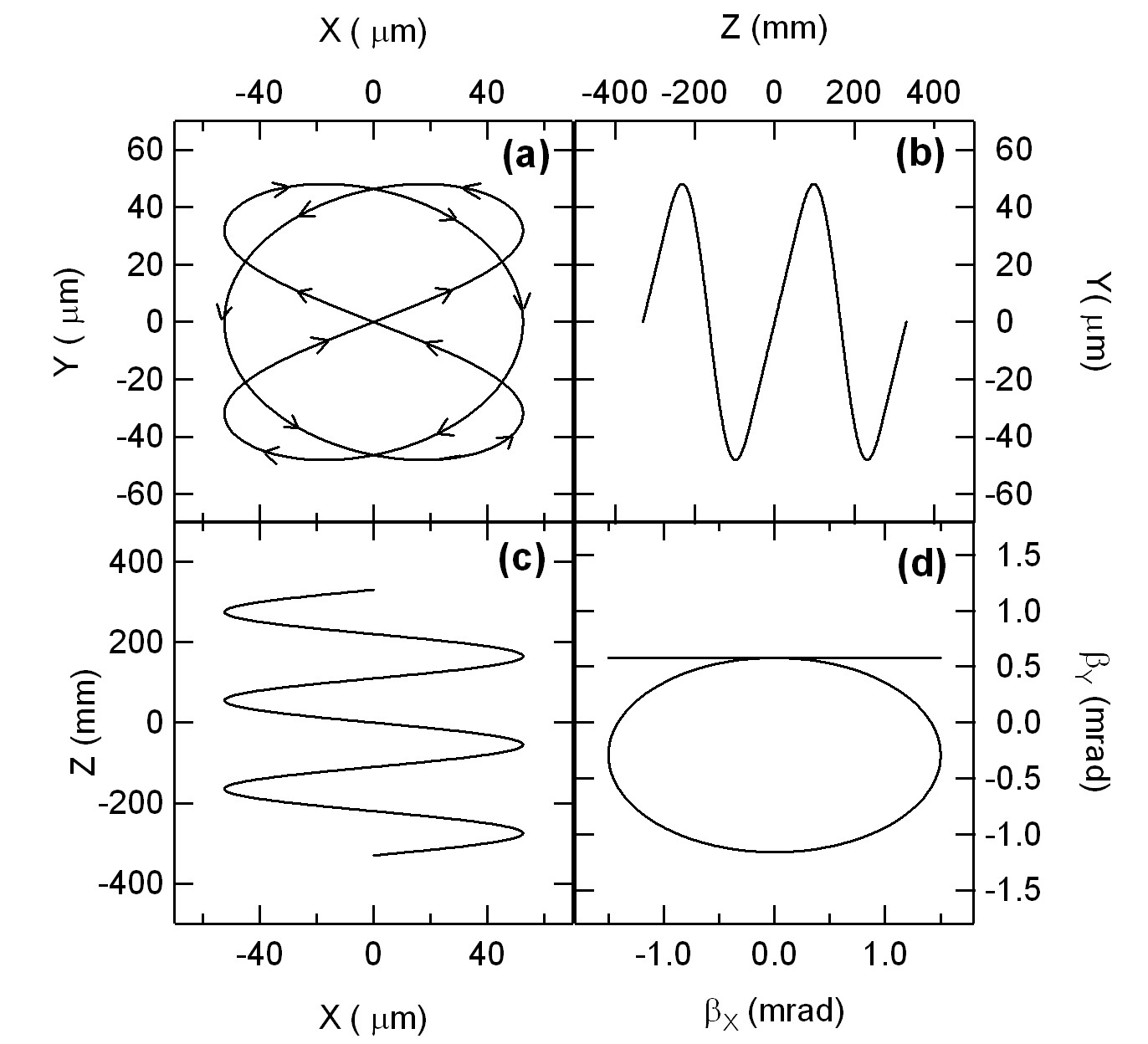}
\caption{The trajectory of electrons in knot undulator projected on the (a) x-y, (b) z-y , (c) x-z planes and (d) the relation between relative velocities in X and Y directions, $\beta_{x}$, $\beta_{y}$. The movement direction of electrons is show by arrows in (a).}
\label{fig3}
\end{figure}

The trajectory and relative velocity of the electrons in knot undulator are show in Fig.\ref{fig3}. The velocity projection on X-Y plan shows successive right- and left-handed elliptical movement and a linear motion (while $B_X$ is zero) between them. We can see that the velocity is always deviated from the undulator axis as we hoped. The projection of trajectory on x-y plan is a complex knot figure. The period of movement in Y direction is 1.5 times of that in X direction, just a reflection of the period ratio of magnetic field in X and Y directions. 

\section{Results and Discussions} 
\begin{figure}[t]
\includegraphics[width=12cm] {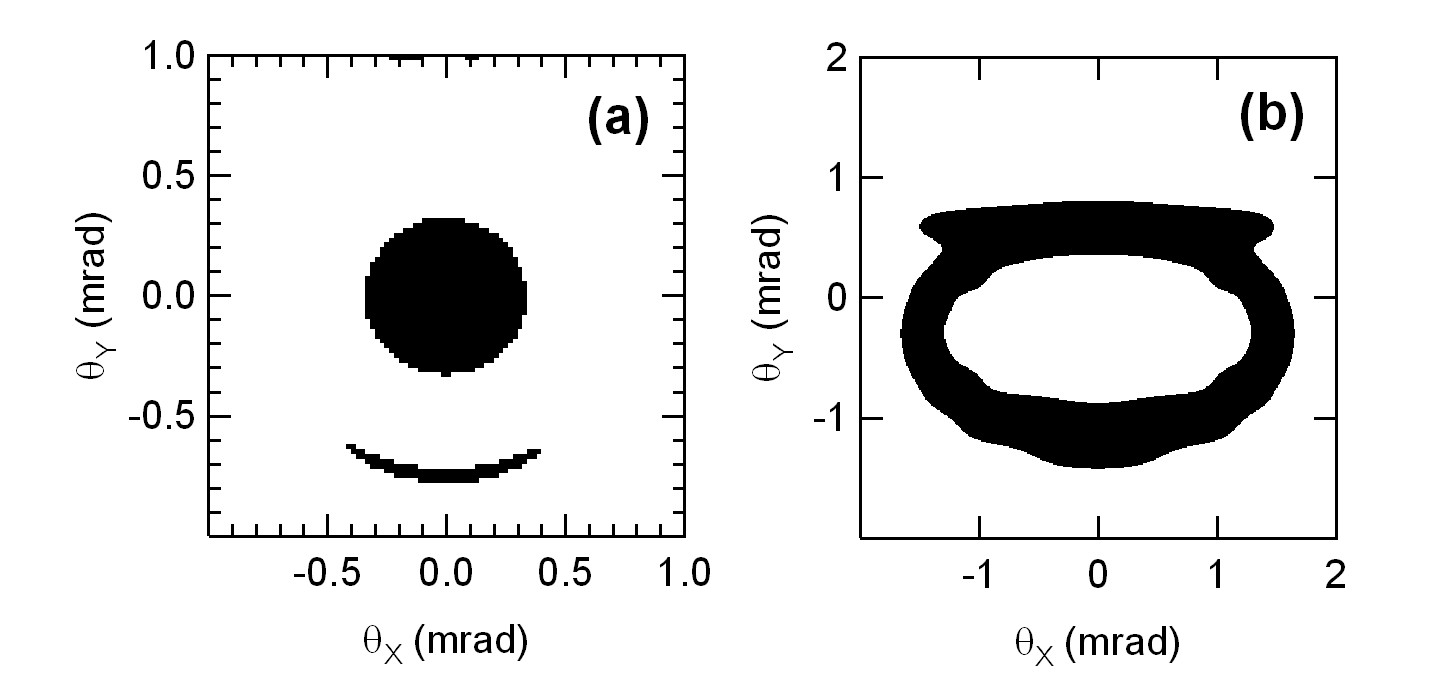}
\caption{The spatial distribution of (a) photon intensity of 7 eV fundamental harmonics and (b) power density.}
\label{fig4}
\end{figure}

\begin{figure}[t]
\includegraphics[width=12cm] {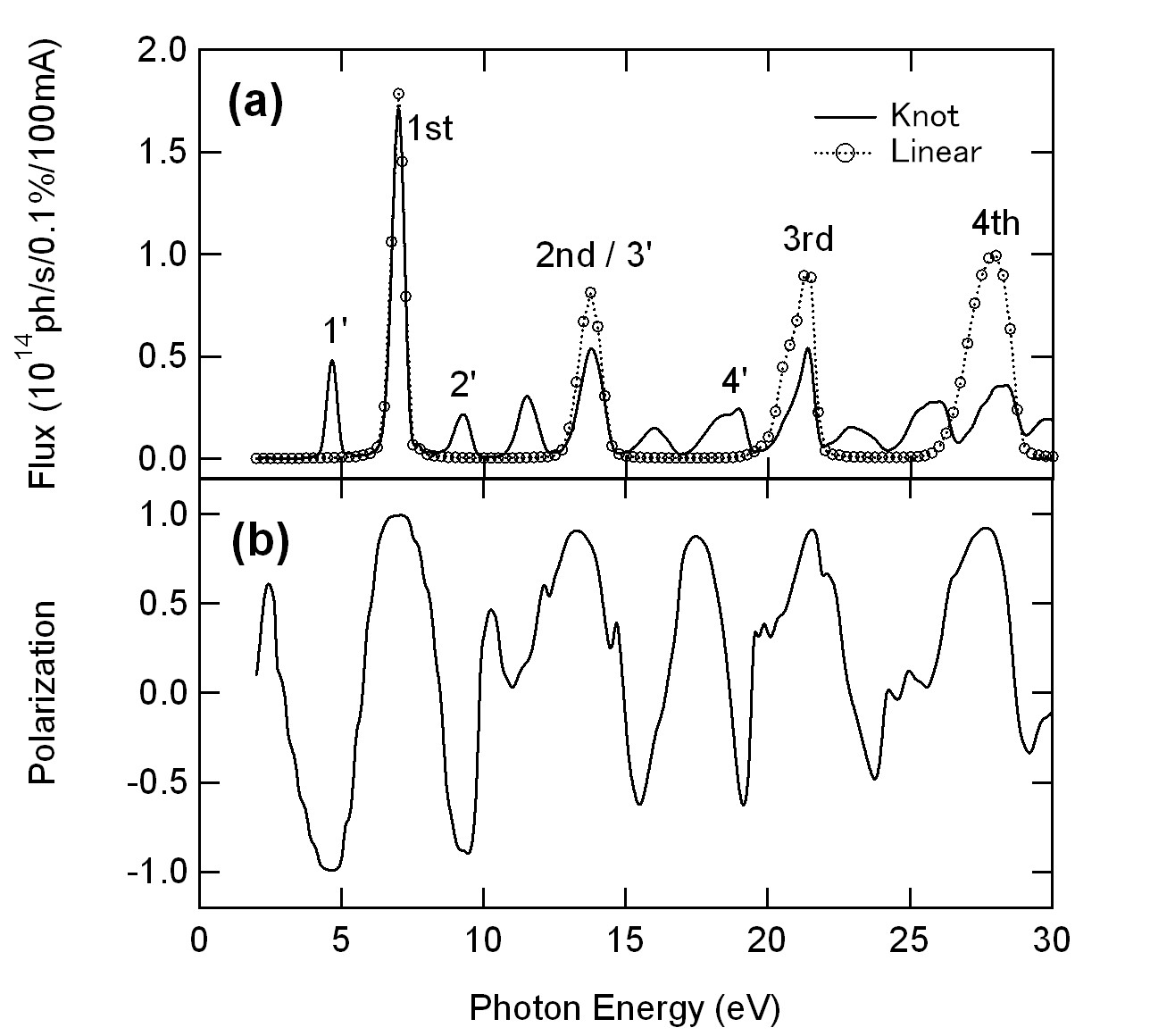}
\caption{(a) The comparison of photon fluxes from knot (solid line) and linear (broken line) undulators. The acceptance angles are 0.6 mrad in both vertical and horizontal directions. (b) The linear polarization of photons from knot undulator within the same acceptance angles as (a).}
\label{fig5}
\end{figure}

All our calculations were carried out with SPECTRA Version 8.05 software developed by T. Tanaka and H. Kitamura of Spring-8 synchrotron radiation facility. For all the calculations, the total length of the undulators is set as 3960 mm (18 periods). The spatial distribution of photon intensity of 7 eV fundamental harmonics is show in Fig.\ref{fig4}(a). We can see that almost all the photons near the undulator axis can be collected if we select acceptance angles of 0.6 mrad in both X and Y directions. The total photon flux and corresponding linear polarization inside this acceptance solid angle is shown in Fig.\ref{fig5} compared with the photon flux from a 18 periods linear undulator with a magnetic field of 0.586 Tesla and a period of 220 mm. We can see that the linear polarization of knot undulator is 99.2\% at the 7eV fundamental peak. We have checked that the spatial intensity distribution of fundamental photons from the linear undulator is almost the same as that from knot one near the undulator axis and the photon flux from the linear undulator is calculated with the same (0.6 mrad, 0.6 mrad) acceptance solid angle. We can see that the photon fluxes from the knot and linear undulators are almost the same. The intensities of high harmonics from knot undulator are smaller than that of linear one, which is the same as the case of elliptical undulators. The flux spectrum from linear undulator shows simple harmonic peaks. The spectrum of knot undulator has three series of peaks. The peaks noted by 1st, 2nd, 3rd and 4th in Fig.\ref{fig5} correspond to the harmonics of 220mm period magnetic field in Y direction  and their polarization are horizontal. While the peaks noted by 1', 2', 3' and 4' are the harmonics of the 330mm period magnetic field in X direction  and their polarization is vertical. The 2nd peak overlaps with the 3E peak, so its polarization is not purely vertical or horizontal. Others are the interference peaks. The spatial distribution of power density of photons from knot undulator is shown in Fig.\ref{fig4}(b). It just corresponds to the velocity directions shown in Fig.\ref{fig3}(d). We can see that most power is out of undulator axis as we expected. The total powers of photons in the (0.6mrad, 0.6mrad) acceptance solid angle from knot and linear undulator are 1.93 and 209 W, respectively, so the heat load from the knot undulator is only 0.92\% of that from the linear one. 

The knot undulator has another very strong advantage. Usually the lowest available photon energy from an electromagnetic EPU is limited by its maximum magnetic field. The weak magnetic field of electromagnetic undulator is a fateful defect in some cases compared to that of permanent magnetic one. By utilizing the knot operating mode, the magnetic field in one direction can help that in another direction to decrease the electron velocity in Z direction, which result in the shift of fundamental energy to the low side, so we can get a lower photon energy compared with that in linear mode. Using knot mode, we can generate a fundamental horizontally polarized 7eV photons by 0.50 and 0.29 T magnetic fields in Y and X directions. Meanwhile the low energy limitation is 9.8 eV for linear mode when only a 0.5 T magnetic field is supplied. This effect is more useful for vertically polarized photons because the horizontal magnetic field is usually weak due to the large gap in this direction. We can examine an existing undulator, UE212 of SIS beamline at Swiss Light Source (SLS), to see the power of knot operation mode. The period, maximum magnetic fields in horizontal and vertical directions of UE212 are 212 mm, 0.1 and 0.4 Tesla, respectively. The beam energy of SLS storage ring is 2.4 GeV. If only linear mode is utilized, the minimum energy for vertical polarized photons is 87 eV due to the 0.1 Tesla weak horizontal magnetic field. By knot mode, the minimum energy of vertically polarized photons can extend to 11.3 eV with a polarization of 97.5\% when 0.1 and 0.4 Tesla magnetic field are supplied in both horizontal and vertical directions. The total available photon flux is eighth of that from a linear undulator, but an unavailable 0.33 Tesla horizontal magnetic field is necessary for linear mode. 

\section{\label{sec4}Conclusions} 
In summary, a novel operating mode for electromagnetic EPU is suggested, which can generate pure linearly polarized photons with very low on-axis heat load. Also the available minimum photon energy of linearly polarized photons can be extended much by this method. 


\end{document}